\documentclass[sigconf]{acmart}
\usepackage{amsmath}
\AtBeginDocument{%
  \providecommand\BibTeX{{%
    \normalfont B\kern-0.5em{\scshape i\kern-0.25em b}\kern-0.8em\TeX}}}

\setcopyright{acmcopyright}
\copyrightyear{2021}
\acmYear{2021}
\acmDOI{10.1145/1122445.1122456}

\acmConference[SIGIR eCom'21]{SIGIR eCom'21: The 2021 SIGIR Workshop On eCommerce}{July 15, 2021}

\setcopyright{rightsretained}
\acmConference[SIGIR eCom'21]{ACM SIGIR Workshop on eCommerce}{July 15, 2021}{Virtual Event, Montreal, Canada}
\acmYear{2021}
\copyrightyear{2021}
\makeatletter
\renewcommand\@formatdoi[1]{\ignorespaces}
\makeatother
\acmISBN{}

\begin{document}

\title{Neural Search: Learning Query and Product Representations in Fashion E-commerce}

 \author{Sagnik Sarkar*}
 \email{sagnik.sarkar@myntra.com}
 \affiliation{%
  \institution{Myntra Designs Pvt. Ltd.}
  \country{India}
 }

 \author{Lakshya Kumar*}
 \email{lakshya.kumar@myntra.com}
 \affiliation{%
  \institution{Myntra Designs Pvt. Ltd.}
  \country{India}
 }
\thanks{*Both authors contributed equally to this research.}
\renewcommand{\shortauthors}{Sagnik and Lakshya, et al.}
\begin{abstract}
Typical e-commerce platforms contain millions of products in the catalog. Users visit these platforms and enter search queries to retrieve their desired products. Therefore, showing the relevant products at the top is essential for the success of e-commerce platforms. We approach this problem by learning low dimension representations for queries and product descriptions by leveraging user click-stream data as our main source of signal for product relevance. Starting from GRU-based architectures as our baseline model, we move towards a more advanced transformer-based architecture. This helps the model to learn contextual representations of queries and products to serve better search results and understand the user intent in an efficient manner. We perform experiments related to pre-training of the Transformer based RoBERTa model using a fashion corpus and fine-tuning it over the triplet loss. Our experiments on the product ranking task show that the RoBERTa model is able to give an improvement of \textbf{7.8\% in Mean Reciprocal Rank(MRR)}, \textbf{15.8\% in Mean Average Precision(MAP)} and \textbf{8.8\% in Normalized Discounted Cumulative Gain(NDCG)}, thus outperforming our GRU based baselines. For the product retrieval task, RoBERTa model is able to outperform other two models with an improvement of \textbf{164.7\% in Precision@50 and 145.3\% in Recall@50}. In order to highlight the importance of pre-training RoBERTa for fashion domain, we qualitatively compare already pre-trained RoBERTa on standard datasets with our custom pre-trained RoBERTa over a fashion corpus for the query token prediction task. Finally, we also show a qualitative comparison between GRU and RoBERTa results for product retrieval task for some test queries. RoBERTa model can be utilized for improving the product search task and act as a good baseline that can be fine-tuned for various information retrieval tasks like query recommendations, query re-formulation, etc.
\end{abstract}



\maketitle

\section{Introduction \& Motivation}
\label{sec:intro and motivation}
Efficient product search in Fashion e-commerce is pivotal for its success. Customers come to the platform with various intents of searching for different fashion products and poor results may cause bad customer experience. Understanding user intent in the form of queries is essential to serve customers with better results. Classic methods of representation learning like LDA\cite{LDA} and LSA\cite{LSA} use unsupervised objectives to map the user query and product in the same vector space for semantic matching. Since the advent of the Word2Vec \cite{mikolov:2013} model, researchers started exploring shallow neural methods for representing queries and products. With several advancements in Natural Language Processing, researchers started experimenting with deeper neural models like LSTM\cite{LSTM}, GRU\cite{GRU}, Transformers\cite{transformers} and BERT\cite{BERT} along with its variants like RoBERTa\cite{Roberta}. For a better understanding of queries and products these deeper models are used to model long-range context and learning better context dependent embeddings. In fact, the Transformer based models have shown to outperform other models for understanding Natural Language by showing significant improvements over standard data sets like GLUE\cite{wang-etal-2018-glue}. When we compare Myntra Fashion E-commerce search with web search engines, the queries that come to our platform are also based on natural language but, they majorly focus on particular products or product categories. Most of the queries are shorter in length which creates a challenge in terms of correctly understanding the user intent and serving relevant results. Some of the example queries are shown below: 
\begin{itemize}
    \item `hrx by hrithik roshan jeans men'
    \item `nike tracksuit men'
    \item `w legging'
    \item `red lehenga choli'
\end{itemize}

In this paper, we propose different neural architectures for understanding queries and products by learning their representation in a low-dimension space. These different models are trained on data that is generated by creating a Query-Product graph and Product-Product co-occurrence graph from Myntra's click-stream data. Our experiments show the comparison of these different models with respect to different ranking metrics on the product ranking task. The performance of these models with respect to the product retrieval task is also discussed. Among the proposed architectures, the transformer-based RoBERTa\cite{Roberta} model is able to achieve the best performance. We show how we have pre-trained an in-house RoBERTa model over a Fashion Corpus that we created using various product descriptions, reviews, and search queries. We also show how this model is fine-tuned for triplet loss optimization. The main contributions of the paper are:
\begin{itemize}
    \item Propose RoBERTa in the task of learning latent query and product representations.
    \item Show a quantitative comparison of RoBERTa and GRU based models on product ranking and retrieval tasks.
    \item Highlight the need for pre-training the RoBERTa model over Fashion Domain.
    \item Report findings of augmenting Product-Product data with Query-Product data while training different neural models.
\end{itemize}
The rest of the paper is organized as follows: In Section \ref{sec:related work}, we review previous works that deal with representing queries/products for tasks like query rewriting, query attribute extraction and product retrieval. In Section \ref{sec:approaches}, we first present a brief overview of data preparation and then present different neural models to learn low-dimensional embeddings of query and product. In Section \ref{sec:experimental setup}, we outline the experimental setup and define different model training strategies. In Section \ref{sec:results}, we present the results and visualizations of our experiments by reporting the performance of proposed models over different tasks. Finally, we conclude the paper and discuss future work in Section \ref{sec:conclusion}.
\section{Related Work}
\label{sec:related work}
Representing queries and products in the same space is useful for tasks like product retrieval, ranking, etc. These tasks are solved in various ways in the e-commerce setting. Traditionally Information Retrieval(IR) in the domain of e-commerce search has been based on exact keyword matching between search queries and product descriptions/titles like BM25\cite{BM25}. But this suffers from the problem of ``vocabulary gap''. With reference to semantic product search in e-commerce, three types of approaches have gained prominence in the recent years, namely, a) Query rewriting, b) Query attribute extraction and c) Embedding-based retrieval.
\subsection{Query rewriting}
The problem of vocabulary gap is pronounced in the e-commerce setting where queries are in informal language whereas the product titles/descriptions are written in formal language. One approach of mitigating this is by re-writing the original query into a query which is semantically similar but has less ``lexical chasm''\cite{learningToRewrite}. The problem can be viewed as a translation task trained on clicked query-document pairs. \cite{TranslationModels} A recent deep learning approach is the ``Learning to Rewrite''\cite{learningToRewrite} framework which leverages the query-product bipartite graph built from click-stream data, to build a candidate query generation phase and a ranking phase to re-write a poorly performing query to a well performing query. Some practical applications of query re-writing in the e-commerce domain include \cite{rewrite_fk}, \cite{rewrite_ebay}.
\subsection{Query attribute extraction}
Typical search queries in e-commerce often include a collection of product attributes that are desired by a customer. One line of work for extracting these attributes from the query is by treating it as a Named Entity Recognition(NER) problem. A weakly-supervised approach for doing this is the work by Guo et al \cite{guo_ner} which applies a weakly supervised LDA algorithm, to identify four types of entities from commercial web search queries containing single named entities. A supervised learning approach is by Cowan et al. \cite{cowan_ner} in the domain of travel search queries, where a linear chain CRF is trained on a manually labeled NER dataset. Some NER based models in e-commerce search are \cite{wen_ner} and \cite{homedepot_ner}. The problem of attribute extraction from search queries can also be treated as a multi-label text categorization problem as demonstrated by Wu et al \cite{google_shopping}
\subsection{Embedding based retrieval}
Classic Embedding Based Retrieval(EBR) methods such as LSA\cite{LSA}, LDA\cite{LDA}, BLTM and DPM\cite{bltm_dpm} involve projecting the query and the document/product into a common embedding space thus capturing concept based similarity. These methods either suffer from being trained on an unsupervised objective which does not align well with the retrieval/ranking task or they have issues with scalability. In recent times, neural network based methods have gained popularity due to their ability of distributed representation learning and scalability. Mitra et. al.\cite{mitra_survey} provides a good survey of neural methods in IR. Guo et. al. \cite{drm_adhoc} categorize deep neural network based methods for IR into 2 categories namely representation-focused models and interaction-focused models. Our approach falls in the category of  representation-focused models since e-commerce product descriptions mostly satisfy the ``Verbosity hypothesis'' \cite{drm_adhoc}. A seminal work in the space of representation-focused models is the DSSM model \cite{dssm} which discriminatively trains a DNN by maximizing the conditional likelihood of the clicked documents given a query using the clickthrough data. New models such as DRMM \cite{drm_adhoc}, Duet \cite{duet} have been further developed to include traditional IR lexical matching signals. Few examples of applying EBR in e-commerce settings include JD.com\cite{Towards_Personalized_and_Semantic_Retrieval} and in other search engines include Search engine EBR papers \cite{fb_ebr} \cite{se_ebr_1}.\\
In this work, we handle the problem of representation learning of queries and products in the same space by proposing different deep models. We report the performance of learned representations from these models on product ranking and retrieval tasks. 

\section{Methodologies}
\label{sec:approaches}
\begin{figure}[h]
  \centering
  \includegraphics[scale = 0.20]{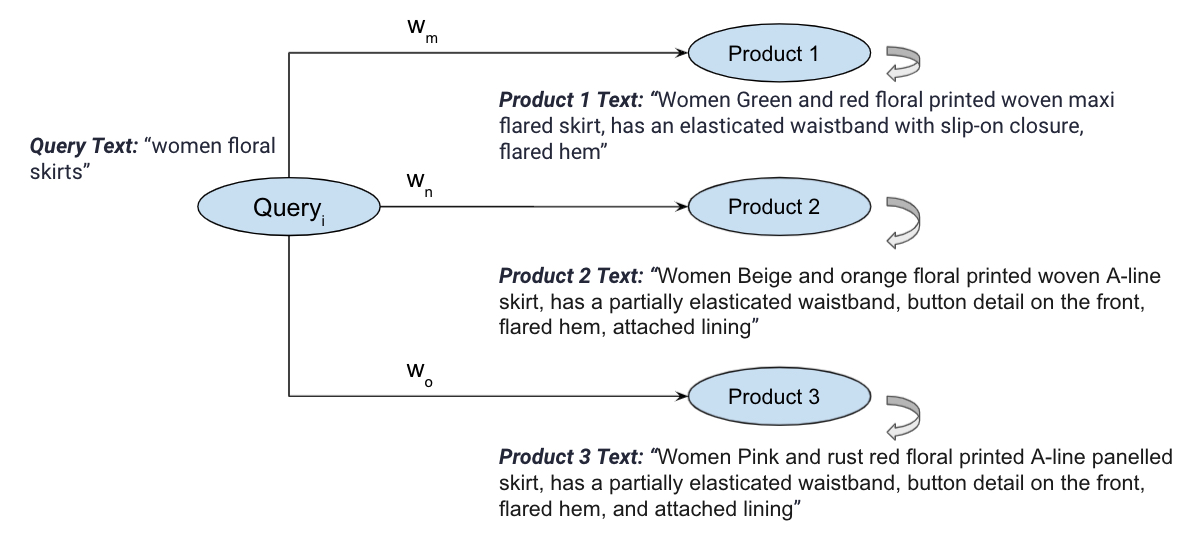}
  \caption{Query-Product Graph}
  \Description{}
  \label{fig:Query Product Graph}
\end{figure}

\begin{figure}[h]
  \centering
  \includegraphics[scale = 0.20]{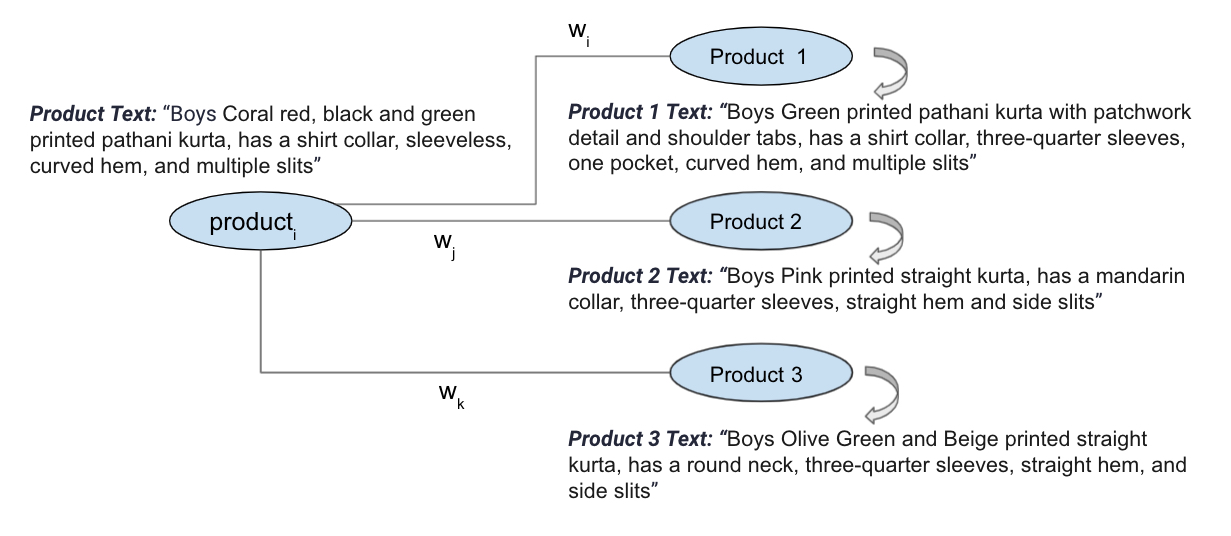}
  \caption{Product-Product Graph}
  \Description{}
  \label{fig:Product Product Graph}
\end{figure}

Myntra's click-stream data is our main source of information for getting products that are relevant to a given query. Using the click-stream data, we form a Query-Product bipartite graph with edge weights representing the number of sessions across which a product is clicked for a given query, as shown in Figure \ref{fig:Query Product Graph} (edge weights represented as $w_{index}$). From the click-stream data, we also form an un-directed Product-Product graph with edge weights representing the number of sessions where 2 products are clicked in the same session, as shown in Figure \ref{fig:Product Product Graph} (edge weights represented as $w_{index}$). 
\subsection{Data Preparation}
\label{sec:Data preparation}

The query-product co-occurrence is mainly driven by the search and ranking engine whereas the product-product co-occurrence is also influenced by the similar product recommendation engine and thus it captures a richer source of information about product-product similarity. All products in Myntra are categorized into various ArticleType-Gender(ATG) groups e.g (AT:Dresses, G:Women), (AT:T-shirts, G:Men) etc. For the Product-Product graph, we only add edges between 2 products belonging to the same ATG. We create (query, positive-product) pairs by taking the query as the anchor and the top 100 products from the Query-Product graph (based on edge-weights) as the positive examples for the query. We observe that clicked results for some queries are limited to 1 ATG whereas for some queries the results span several ATGs. We mark queries spanning several ATGs as ``unnamed queries'' since they do not refer to a specific ATG. For example, the query ``bags'' would result in different types of bags such as \textit{ladies handbags, laptop bags, trolleys} etc. We remove such queries from our dataset as they are already handled well by the current search engine. For queries spanning a single ATG, we observe that some queries are very broad in nature and almost all products in the relevant ATG may be relevant to the query for e.g: ``men tshirts'' (ATG: (T-shirts, Men)). Any query whose clicked results cover more than 30$\%$ of the products in the relevant ATG is marked as a ``broad query'' and the rest are marked as ``narrow queries''. As we use triplet loss in order to optimize different neural models, we follow 2 types of negative sampling strategies for mining negative examples to calculate the triplet loss. For ``broad queries'', we randomly sample a product from a different ATG (different from that of the positive product) whereas for ``narrow queries'' such as \textit{``skinny fit jeans for men''}, half of the time, we randomly sample a product from the same ATG as that of the positive product and otherwise we sample from a different ATG. We also use the Product-Product  graph to generate anchor-positive pairs of products. This is done by simulating a fixed number of fixed length short random walks from each product node (anchor) and making all visited nodes its positive examples. The negative products are  randomly sampled half of the time from the same ATG as the anchor and rest of the times from a different ATG. These product-product pairs are also augmented to the query-product pairs for training in one set of experiments. Using the mentioned approach, we finally obtain two sets of data, i.e., Query-Product and Product-Product data. We explain different neural models that we develop in order to learn the latent representation for queries and products in Myntra Fashion e-commerce. The basic entity for the first two neural architecture is GRU cell\cite{GRU}. The third neural model is based on Transformer\cite{transformers}, i.e., RoBERTa\cite{Roberta} model which is first pre-trained on the Fashion corpus that we created manually and then fine-tuned by optimizing Triplet loss. We experiment with each of these different deep models in two settings, i.e., first by training them on Query-Product data and then by augmenting Query-Product data with Product-Product data for training.

\subsection{Gated Recurrent Units(GRU)}
\label{sec:GRU cell}
The GRU has gating units like LSTM\cite{LSTM} that control the flow of information inside the unit. The GRU is different from LSTM as it does not have separate memory cells. The GRU cell mainly consists of two gates, i.e., Reset Gate and Update Gate. These gates are described below\footnote{The definition of GRU cell and equations are taken from https://d2l.ai/chapter\_recurrent-modern/gru.html}: 
\begin{itemize}
    \item \textbf{Reset Gate}: Helps to control how much of the previous state information will be remembered.
    \item \textbf{Update Gate}: Controls the amount of previous information to throw away and what new information to add based on the current input to the GRU unit. This gate acts similar to the forget and input gate present in LSTM cell unit. 
\end{itemize}
Formally, for a given time step t, assume that the input minibatch $X_{t} \in \mathbb{R}^{nxd}$ with number of examples as n and the number of inputs as d. Also, the hidden state of the previous time step is $H_{t-1} \in \mathbb{R}^{nxh}$ with number of hidden units as h. Then, the update gate $Z_{t} \in \mathbb{R}^{nxh}$ and reset gate $R_{t} \in \mathbb{R}^{nxh}$ are obtained as follows:
\begin{equation}
    R_{t} = \sigma({X_{t}W_{xr} + H_{t-1}W_{hr} + b_{r}})
\end{equation}
\begin{equation}
    Z_{t} = \sigma({X_{t}W_{xz} + H_{t-1}W_{hz} + b_{z}})
\end{equation}
where $W_{xr}, W_{xz} \in \mathbb{R}^{dxh}$ and $W_{hr}, W_{hz} \in \mathbb{R}^{hxh}$ are weights and $b_{r}, b_{z} \in \mathbb{R}^{1xh}$ are biases. The candidate hidden state $\Tilde{H}_{t} \in \mathbb{R}^{nxh}$ at time step t,
\begin{equation}
     \Tilde{H}_{t} = \tanh({X_{t}W_{xh} + (R_{t}\odot H_{t-1})W_{hh} + b_{h}})
\end{equation}
where $W_{xh} \in \mathbb{R}^{dxh}$ and $W_{hh} \in \mathbb{R}^{hxh} $ are weights and $b_{h} \in \mathbb{R}^{1xh}$ is bias.
The equation that leads to hidden state at time step t is given as,
\begin{equation}
    {H}_{t} =   Z_{t}\odot H_{t-1} + (1-Z_{t})\odot \Tilde{H}_{t}
\end{equation}
GRU has fewer tensor operations as compared to LSTM and they are faster to train as well, so we choose GRU cell unit in order to build first two neural architectures. Below we describe bi-directional GRU based neural model, where first neural model has only single layer, i.e., one forward GRU and one backward GRU in order to read the textual sentence. The second neural model is more deep as it consists of two layers, i.e., two forward GRU and two backward GRU. 

\subsection{BiGRU: Single Layer}
\label{sec:BiGRU: Single Layer}

\begin{figure}[h]
  \centering
  \includegraphics[scale = 0.21]{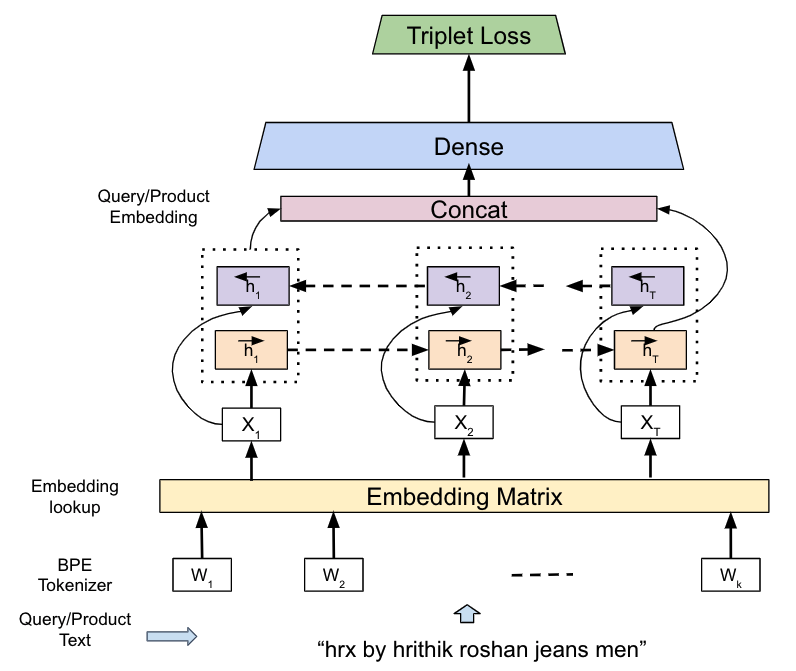}
  \caption{BiGRU:Single Layer Architecture}
  \Description{}
  \label{fig:Single Layer architecture BiGRU}
\end{figure}

Single Layer BiGRU based model is shown in Figure \ref{fig:Single Layer architecture BiGRU}. It consists of one forward and one backward GRU layer in order to learn the query/product representations in latent space. We train a BPE\cite{BPE} tokenizer over the Fashion corpus\footnote{Consists of queries, product titles, descriptions and reviews etc.} and use it in all the proposed neural models for tokenizing the input. The input to the model can be a query text or product description. As shown in the architecture, the last forward and backward GRU hidden states are concatenated and passed to a fully connected layer and finally the model is trained to optimize for the Triplet Loss function. The mathematical formulation of the Triplet Loss function is given as: 
\begin{equation}
    Loss = \sum_{i=1}^{N}[d(f_{i}^{a},f_{i}^{p}) - d(f_{i}^{a},f_{i}^{n}) +\alpha]_{+}
    \label{eq:Triplet Loss}
\end{equation}
where $f_{i}^{a}$, $f_{i}^{p}$, $f_{i}^{n}$ correspond to the embeddings of query/product, positive product(clicked product) and negative products(as mentioned in Section \ref{sec:Data preparation})  obtained from the model respectively. And $d$ denotes the distance metric which is cosine distance in our model optimization. $[]_{+}$ denotes the function which will be non-zero if the value inside is positive and zero otherwise. The model optimization will try to bring the query/product embedding closer to positive product, i.e., clicked product and it will push away the random negative products from the query/product embedding. We have used this loss function in all our neural models in order to do the model training and then obtained the embeddings from the trained model for doing the evaluation over test dataset.

\subsection{BiGRU: Multi Layer}
\label{sec:BiGRU: Multi Layer}

\begin{figure}[h]
  \centering
  \includegraphics[scale = 0.25]{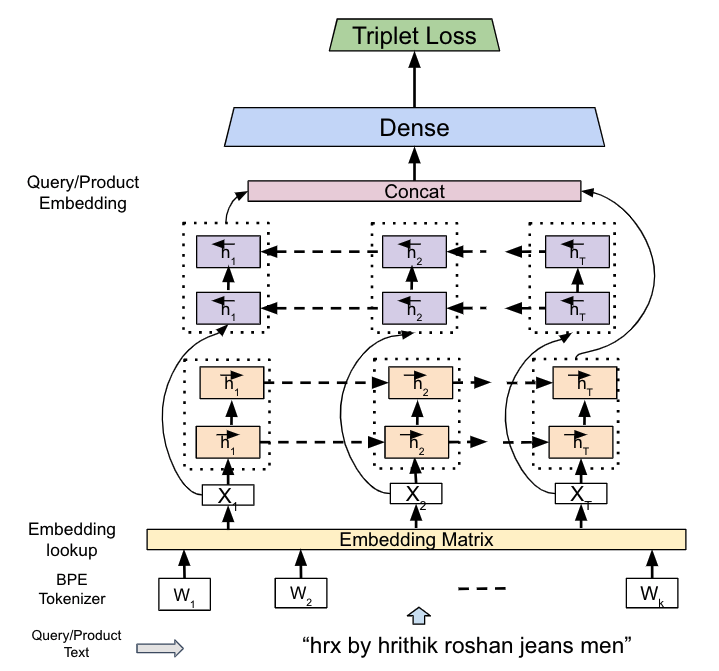}
  \caption{BiGRU:Multi Layer Architecture}
  \Description{}
  \label{fig:Multi Layer architecture BiGRU}
\end{figure}

In order to increase the model capacity, we have introduced another BiGRU layer and the same is shown in Figure \ref{fig:Multi Layer architecture BiGRU}. This architecture contains two forward GRU Layers and two backward GRU layers. The BPE tokenizer that is shown is same as described in Section \ref{sec:BiGRU: Single Layer}. This model is almost similar to single layer BiGRU architecture proposed in Figure \ref{fig:Single Layer architecture BiGRU} except for one more layer that gives the model more flexibility to learn better query/product representations that result in improved metrics as discussed in Section \ref{sec:results}. This model is also trained by optimizing for Triplet Loss function which is given in Equation \ref{eq:Triplet Loss}. Each of the proposed neural model is capable to take query text or product text as input. In Section \ref{sec:experimental setup}, we will describe how we train each of the proposed model by giving two types of data, i.e., Query-Product data(obtained from Query-Product graph) and Product-Product data(which is obtained from Product-Product co-occurence graph).

\subsection{RoBERTa model}
\label{sec: RoBERTa Approach}
\begin{figure}[h]
  \centering
  \includegraphics[scale = 0.22]{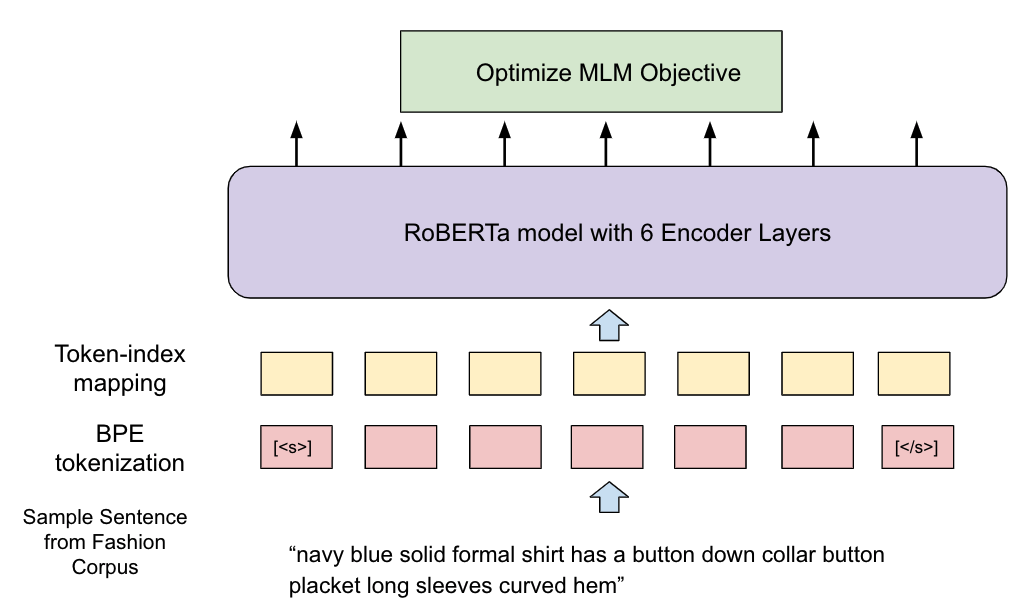}
  \caption{RoBERTa Pre-training setup using Fashion Corpus}
  \Description{}
  \label{fig:BERT Model Pretraining setup}
\end{figure}

RoBERTa\cite{Roberta} model which is a variant of BERT\cite{BERT} is used for learning Fashion language using Fashion Corpus that is described in Section \ref{sec:experimental setup}. 
The BERT model optimizes over two auxiliary pre-training tasks: 
 \begin{itemize}
     \item \textbf{Mask Language Model (MLM)}: Randomly masking 15\% of the tokens in each sequence and predicting the masked tokens
     \item \textbf{Next Sentence Prediction (NSP)}: Randomly sampling sentence pairs and predicting whether the latter sentence is the next sentence of the former
 \end{itemize}
 BERT based representations try to learn the context around a word and is able to better capture its meaning syntactically and semantically. For our case, we directly use RoBERTa model as it only optimises for MLM auxiliary task which is sufficient for efficient pre-training as shown in \cite{Roberta}. In experiments we use byte-level BPE \cite{BPE} tokenization\footnote{The tokenizer is same across both the GRU and RoBERTa based models.} for encoding sentences present in Fashion corpus. We use perplexity\cite{chen_beeferman_rosenfeld_2018} score for evaluating the RoBERTa language model. Perplexity is defined as the exponentiated average log-likelihood of a sequence. If we have a tokenized sequence $X = (x_{0}, x_{1}, x_{2},...., x_{t} )$ then perplexity of X is, 
 \begin{gather}
    	PPL(X) = \exp{\{ \frac{-1}{t}\sum_{i}^{t}\log P_{\theta}(x_{i}|x_{<i})\}}     
\end{gather}
where $\log P_{\theta}(x_{i}|x_{<i})$ is the log-likelihood of the ith token conditioned on the preceding tokens $x_{<i}$ according to RoBERTa model, where $\theta$ indicates the model parameters.  Generally, lower is the perplexity, better is the language model. After pre-training, the model would have learnt the syntactic and semantic aspects of tokens present in sentences in the Fashion corpus. With the help of self-attention it learns the context in which different tokens appear and tries to predict masked tokens based on the left and right context. Masking helps the RoBERTa model to use both the left and the right context without facing the problem of data leakage and the model learns contextual low dimension representation of tokens. The pre-training setup of RoBERTa model is shown in Figure \ref{fig:BERT Model Pretraining setup}. Again the BPE tokenizer used is same as mentioned in other neural approaches described in Sections \ref{sec:BiGRU: Single Layer} and \ref{sec:BiGRU: Multi Layer}. 

\begin{figure}[h]
  \centering
  \includegraphics[scale = 0.22]{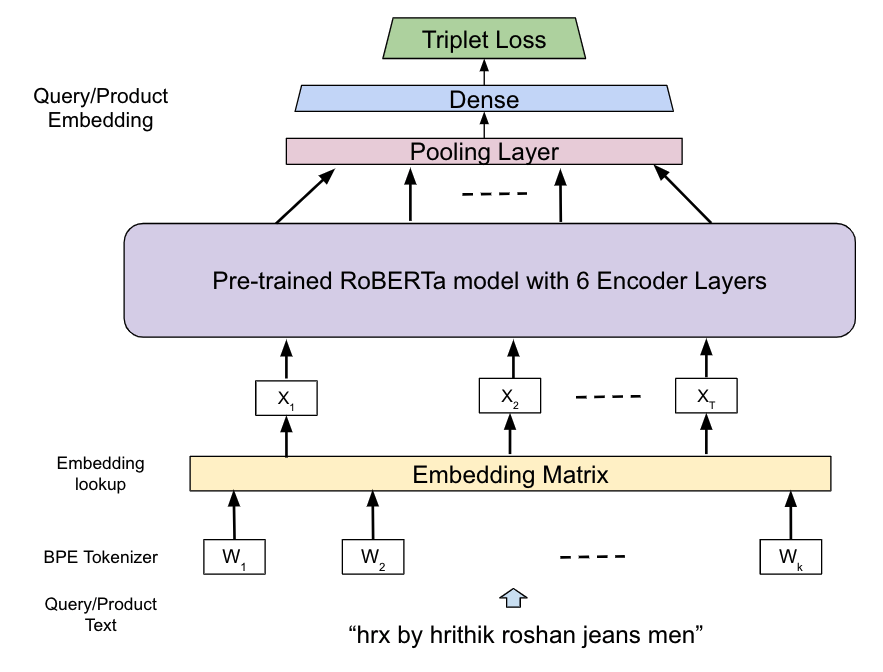}
  \caption{Fine-Tuning RoBERTa model with Triplet Loss}
  \Description{}
  \label{fig:BERT Model Fine-tuning}
\end{figure}
After Pre-training the RoBERTa model over Fashion corpus, we fine tune it to optimize for the Triplet loss function. The architecture for fine-tuning the pre-trained RoBERTa model is shown in Figure \ref{fig:BERT Model Fine-tuning}. Similar to other BiGRU based neural models, we fine-tune it over Query-Product and Product-Product data as described in Section \ref{sec:experimental setup}. A query text or product description can be given as input to this model after tokenizing through BPE tokenizer and then the model will generate the embeddings corresponding to different tokens. A pooling layer is applied to the embeddings to generate a single embedding which acts as a latent representation of the input. Again the Triplet Loss is optimized using this model in order to reduce the distance between query/product and clicked product and increase the distance between query/product and random negatively sampled products.\\
The three neural approaches mentioned are compared by computing different metrics like Mean Reciprocal Rank(MRR), Mean Average Precision(MAP) and Normalized Discounted Cumulative Gain(NDCG) for product ranking task. We also evaluate the proposed models for product retrieval task by calculating precision@K and recall@K metrics. The quantitative analysis is explained in Section \ref{sec:results}. 

\section{Experimental Setup}
\label{sec:experimental setup}
\subsection{Dataset Description}
For all our experiments we use 60 days of click-stream data. Spark\cite{spark} framework is used to prepare the Query-Product and Product-Product graphs. The resulting data contains approximately \textbf{140k} ``unique''(in terms of exact matching) queries and \textbf{950k} products. We randomly split the queries into train and test sets in the ratio of 85:15 resulting in around \textbf{119k} train queries and \textbf{21k} test queries. We only use queries in the train set to form the query-product pairs. The Product-Product graph is also formed from click-stream events in the same time-span as the Query-Product graph. For generating product-product pairs for data augmentation, we simulate 5 random walks of length 5 per product node in the product co-occurrence graph and also remove repeating nodes from the random walk. We use the igraph\cite{igraph} library for simulating the random walks.

\subsection{Model Training}
We train different models proposed in Section \ref{sec:approaches} in different data settings. Each of the mentioned neural model is trained using only Query-Product data as well as a larger data consisting of Query-Product and Product-Product data. We call the models trained with the larger data as models with Augmented Data as shown in Table \ref{table:Query-click product prediction task} and \ref{table:Product Retrieval Task}. We will first describe the settings and the framework used to train GRU based neural models and then explain the pre-training and fine-tuning of the RoBERTa model. For all the models, we train the Byte-Pair Encoding(BPE) tokenizer over the whole dataset consisting of product descriptions, queries and product reviews. This dataset is called as Fashion Corpus. This tokenizer is used in order to tokenize the model input. The vocabulary size of the tokenizer is kept as \textbf{30K}. The Query-Product training data consists of query along with one clicked product that acts as positive and a randomly sampled product that acts as negative. The negative product is sampled from the same ATG or from a different ATG based on ``narrow'' or ``broad'' query type as mentioned in Section \ref{sec:Data preparation}. For example: If the query is \textit{`women kurtas'}, the positive product is the one which is clicked by the user and the negative product is sampled from different ATG because it is a broad query. In each of the neural models, the triplet loss optimization is performed by taking these positive and negative examples. For Product-Product data, we have an anchor product and other positive product which co-occurred with the anchor product in the random walk over Product-Product Graph. For this data, the negative product is sampled half of the time from the same ATG and half of the time from a different ATG. The model architectures proposed in Section \ref{sec:approaches} are trained first on Query-Product data and then by augmenting Product-Product data to see the impact of adding the second data on model performance. The results with respect to different metrics are explained in Section \ref{sec:results}.
\subsubsection{\textbf{GRU based Neural Models}}
The single layer BiGRU model is trained in Pytorch\cite{pytorch} Deep Learning framework. The Embedding layer of the model is $E \in \mathbb{R}^{|V| x D}$, where $|V|$ is vocabulory size which is 30K and $D$ is the input token embedding dimension which is  \textbf{100}. The hidden unit size of GRU cell is also \textbf{100}. There is a dense layer after the BiGRU as shown in Figure \ref{fig:Single Layer architecture BiGRU}. After the forward and backward GRU cells read the tokenized input, their last hidden states are concatenated and given as input to dense layer with $W_{Dense} \in \mathbb{R}^{200x100}$. While training the model, we pass query text as well as product description for both the positive product and negatively sampled product one after another and generate their embeddings. Finally, the Triplet loss optimization is performed over these embeddings to minimze the loss. The multi-layer BiGRU neural model which is shown in Figure \ref{fig:Multi Layer architecture BiGRU} is trained in a similar fashion. The number of BiGRU layers in this model are 2, i.e., two forward GRU and two backward GRU cells. After taking the tokenized input, forward GRU reads it and finally generates the hidden state from second GRU cell. In a similar way, the second GRU cell of backward GRU produce the final state. These two hidden states are concatenated and passed to the dense layer as shown in the architecture. All the other hyperparameters for this model are kept same except from two BiGRU layers instead of one. GRU based neural models uses Adam\cite{Adam} optimizer in order to update the parameters of the model. All the GRU based models are trained for \textbf{50 epochs}.

\subsubsection{\textbf{RoBERTa based Neural Models}}
In order to model the problem of query/product representation learning, we first train the RoBERTa model from scratch to learn the different syntactic and semantic aspects of words appearing in the context of Fashion. We have prepared a Fashion corpus of size \textbf{$\sim4.5GB$} consisting of product titles, descriptions, product reviews and queries. This corpus is used to pre-train the RoBERTa language model and then the pre-trained RoBERTa model is fine-tuned for the Triplet loss optimization.  
\paragraph{\textbf{Pre-training over Fashion Corpus}}
For RoBERTa model pre-training, the architecture is shown in Figure \ref{fig:BERT Model Pretraining setup}.
The model is trained to optimize `\textbf{Masked Language Modelling}' objective as mentioned in \ref{sec: RoBERTa Approach} for \textbf{2 epochs} with a cumulative training time of \textbf{2.5 days} and per-gpu training batch size of \textbf{8}. The multi-gpu pre-training is done using \textbf{Pytorch}\cite{NEURIPS2019_9015} framework and \textbf{HuggingFace} library\cite{wolf2019huggingfaces} based implementation of RoBERTa model with 2 \textbf{Tesla V100 GPUs}. The architecture of RoBERTa model that is used is \textbf{‘DistilRoBERTa- base’} from HuggingFace having \textbf{6 encoder layers, 12 attention heads per layer and 82 million parameters} and this model is called as \textit{RoBERTaForMaskedLM}. The hidden embedding dimension of the RoBERTa model is \textbf{768} and the position embedding is also of the same dimension. The model accepts the tokenized input of maximum length \textbf{512}. The model uses \textbf{AdamW} \cite{kingma2014method}\cite{loshchilov2017decoupled} optimizer in order to update the parameters during training. The perplexity of the RoBERTa model over evaluation dataset is obtained as \textbf{3.5}. In Section \ref{sec:results}, we show how the RoBERTa model pre-trained over Fashion corpus is able to capture the context and predict the masked words with relevant words as compared to already pre-trained RoBERTa model that is pre-trained on standard datasets.
\paragraph{\textbf{Fine-tuning RoBERTa model}}
The fine tuning architecture of the RoBERTa model is shown in Figure \ref{fig:BERT Model Fine-tuning}. After pre-training the RoBERTa model from scratch on Fashion corpus it is fine-tuned for the triplet loss optimization. In order to prevent deviation of model parameters while fine-tuning, we use slanted triangular learning rate\cite{ULMFit} strategy which first linearly increases the learning rate and then linearly decays it as per defined update rule. In order to perform this fine-tuning, we use scheduler present in the HuggingFace library. The optimizer for updating the parameters of this model is same as mentioned in pre-training of the RoBERTa model. The architecture for this model is same as the pre-trained RoBERTa model except that there is one RoBERTa pooling layer present at the top of the model. The class name of the model is called \textit{RoBERTaModel} in the HuggingFace. When we initialize the \textit{RoBERTaModel} with the pre-trained model, all the weights get initialized with the pre-trained weights. Given the tokenized input text, this model will output a embedding of dimension \textbf{768} which is then passed to a dense layer of dimension:$W_{Dense} \in \mathbb{R}^{768x100}$. Finally, we optimize this model for the Triplet loss function in order to fine-tune the model in two different data settings, i.e., training on Query-Product data and training by augmenting Product-Product data. This model is fine-tuned for only \textbf{1 epoch}.\\ 
We report the performance of different models for ranking and retrieval tasks in Section \ref{sec:results}. Among different models, RoBERTa model act as a good baseline for serving the relevant products over fashion e-commerce given the user query.

\section{Results \& Visualization}
\label{sec:results}
In order to evaluate the embeddings obtained from different models proposed in Section \ref{sec:approaches}, we choose two tasks: Product Ranking and Product Retrieval. For these two tasks, we report different metrics.

\subsection{Product Ranking Task}
In order to compare different neural models with respect to a downstream task, we use query-clicked product ranking task. Table \ref{table:Query-click product prediction task} show different ranking metrics computed for different models. In order to compute Mean Reciprocal Rank(MRR) for all the models with respect to clicked product, we have created a test dataset containing one clicked product and 20 negative products for every query, so the ratio of positive to negative is 1:20. All the models are evaluated on this test dataset. Among the different neural baselines, the single layer BiGRU with augmented data is able to give \textbf{38.8\%} MRR. BiGRU with 2 layers trained over only Query-Product data give a MRR score of \textbf{44.5\%}. The fine-tuned RoBERTa model give a MRR score of \textbf{48\%} which is trained with only Query-Product data and a MRR of \textbf{40\%} with augmenting Product-Product data. For Multi-layer GRU, we observe a performance drop in the MRR but an increase in MAP and NDCG with augmented data. For the RoBERTa model, we observe a decline in all the ranking metrics with augmented data. We suspect that the drop is due to fine-tuning only for 1 epoch even after augmenting more data. The RoBERTa model is able to outperform both the single layer GRU and multi layer GRU with just 1 epoch of fine-tuning as compared to 50 epochs for the other models. As RoBERTa model is already pre-trained over Fashion corpus, it is able to achieve a higher MRR score thereby leveraging transfer learning. In order to further compare these different models, we compute other ranking metrics like MAP and NDCG. We prepare another smaller test dataset from Query-Product graph containing positive(clicked) products and 3 random negative products per positive product with respect to each query. In this smaller dataset, the queries are a subset of bigger dataset that we use for computing MRR metric. For every query in this smaller dataset, we rank the set of positive and negative products using the embeddings obtained from the trained models. For example, if for a given query there are $n_{1}$ positive(clicked) products and $n_{2}=3n_{1}$ negative products, then we first obtain the embedding for query text. Then we obtain the product embedding for each of the $n_{1} + n_{2}$ products using their product description from the trained models. All the $n_{1} + n_{2}$ products are ranked using cosine-similarity between query embedding and product embedding. After generating the predicted ranking, we compare it with the ideal ranking where all the positive(clicked) product should present before all the negative products(non-clicked).
With respect to this dataset, the RoBERTa model fine-tuned with only Query-Product data outperforms other models with a MAP of \textbf{70.9\%} and NDCG of \textbf{81.1\%}.

\begin{table}[!h]
\centering
  \resizebox{180 pt}{!}{%
   \begin{tabular}{|c|c|c|c|}\hline
     \textbf{Approach} & \textbf{MRR} & \textbf{MAP} &\textbf{NDCG} \\ \hline
   
    GRU:Single Layer   & 29.36{\%} & 59.7{\%} & 73.3{\%}\\ \hline 
    
    GRU:Single Layer with Augmented Data  & 38.8{\%} & 60.9{\%} & 74.2{\%}\\ \hline 
    
    GRU:Multi Layer & 44.5{\%} & 58.9{\%} & 72.7{\%}\\ \hline
    
    GRU:Multi Layer with Augmented Data & 41.5{\%} & 61.2{\%} & 74.5{\%}\\ \hline 
    
    RoBERTa model & \textbf{48{\%}} & \textbf{70.9{\%}} & \textbf{81.1{\%}}\\ \hline 
    
    RoBERTa model with Augmented Data & 40{\%} & 63.6{\%} & 76.1{\%}\\ \hline 
    
  \end{tabular}} 
\caption{Evaluation for product ranking task}
\label{table:Query-click product prediction task}
\vspace{-6mm}
\end{table}

\subsection{Product Retrieval Task}
The query and product embeddings learned using different models can be used to retrieve the products given the query embedding. A comparison of all the models with respect to product retrieval task is shown in Table \ref{table:Product Retrieval Task}. For product retrieval task, we use the same dataset that we use to calculate MAP and NDCG and filtered only those queries that belong to 1 ATG. In order to assign the ATG to queries, we look at the clicked product for the queries and assign the ATG of the clicked products to the queries.\\ 
For retrieving the products based on query embedding, we construct six\footnote{which corresponds to 6 different models trained as shown in Table \ref{table:Query-click product prediction task} and Table \ref{table:Product Retrieval Task}.} different Annoy\cite{Annoy} index over the product embeddings obtained from different models. For every query in the test dataset, we obtain the query embedding from the model and then refer the Annoy index to fetch top 50 products. Along with each query, we also have a ground truth set of clicked products obtained from Query-Product graph that is used to compute the Precision@50 and Recall@50 metrics. Among the different models, RoBERTa model outperform all other models with a Precision@50 of \textbf{4.5\%} and Recall@50 of \textbf{21.1\%}. 
\begin{table}[!h]
\centering
  \resizebox{180 pt}{!}{%
   \begin{tabular}{|c|c|c|c|}\hline
     \textbf{Approach} & \textbf{Precision@50 \%} & \textbf{Recall@50 \%}  \\ \hline
   
    GRU:Single Layer   & 1.4\% & 7.5{\%} \\ \hline 
    
    GRU:Single Layer with Augmented Data  & 1.7\% & 8.6{\%} \\ \hline 
    
    GRU:Multi Layer & 1.4\%  & 7.4{\%} \\ \hline
    
    GRU:Multi Layer with Augmented Data & 1.2\% & 6.1{\%}\\ \hline 
    
    RoBERTa model & \textbf{4.5\%} & \textbf{21.1{\%}}\\ \hline 
    
    RoBERTa model with Augmented Data & 2.7\% & 11.1{\%}\\ \hline 
    
  \end{tabular}} 
\caption{Evaluation for product retrieval task}
\label{table:Product Retrieval Task}
\vspace{-6mm}
\end{table}

\begin{figure}[!h]
  \centering
  \includegraphics[scale = 0.10]{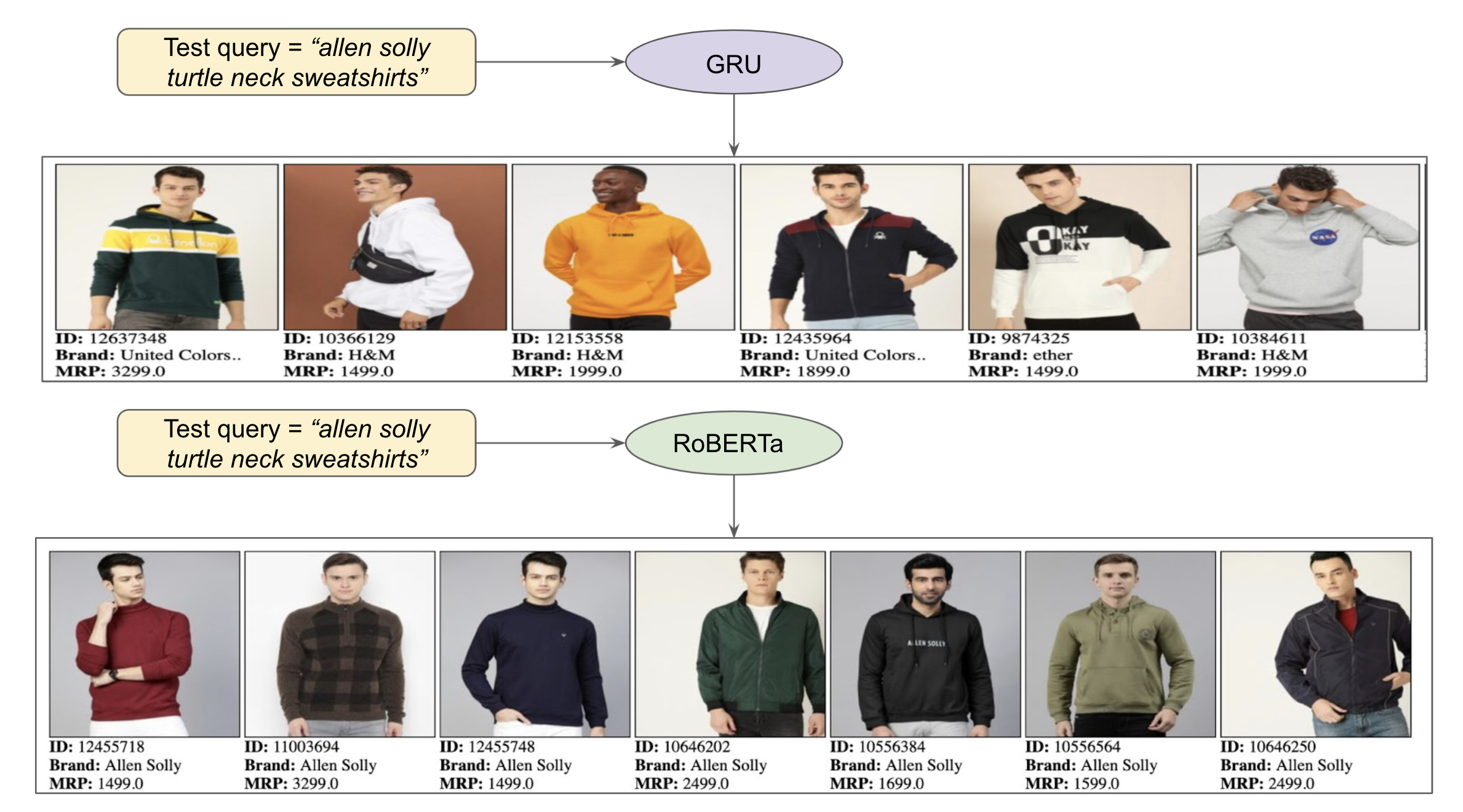}
  \caption{Comparison of retrieved product results for test query: \textit{`allen solly turtle neck sweatshirts'}}
  \Description{}
  \label{fig:ranking_visualization-2}
\end{figure}

\begin{figure}[!h]
  \centering
  \includegraphics[scale = 0.10]{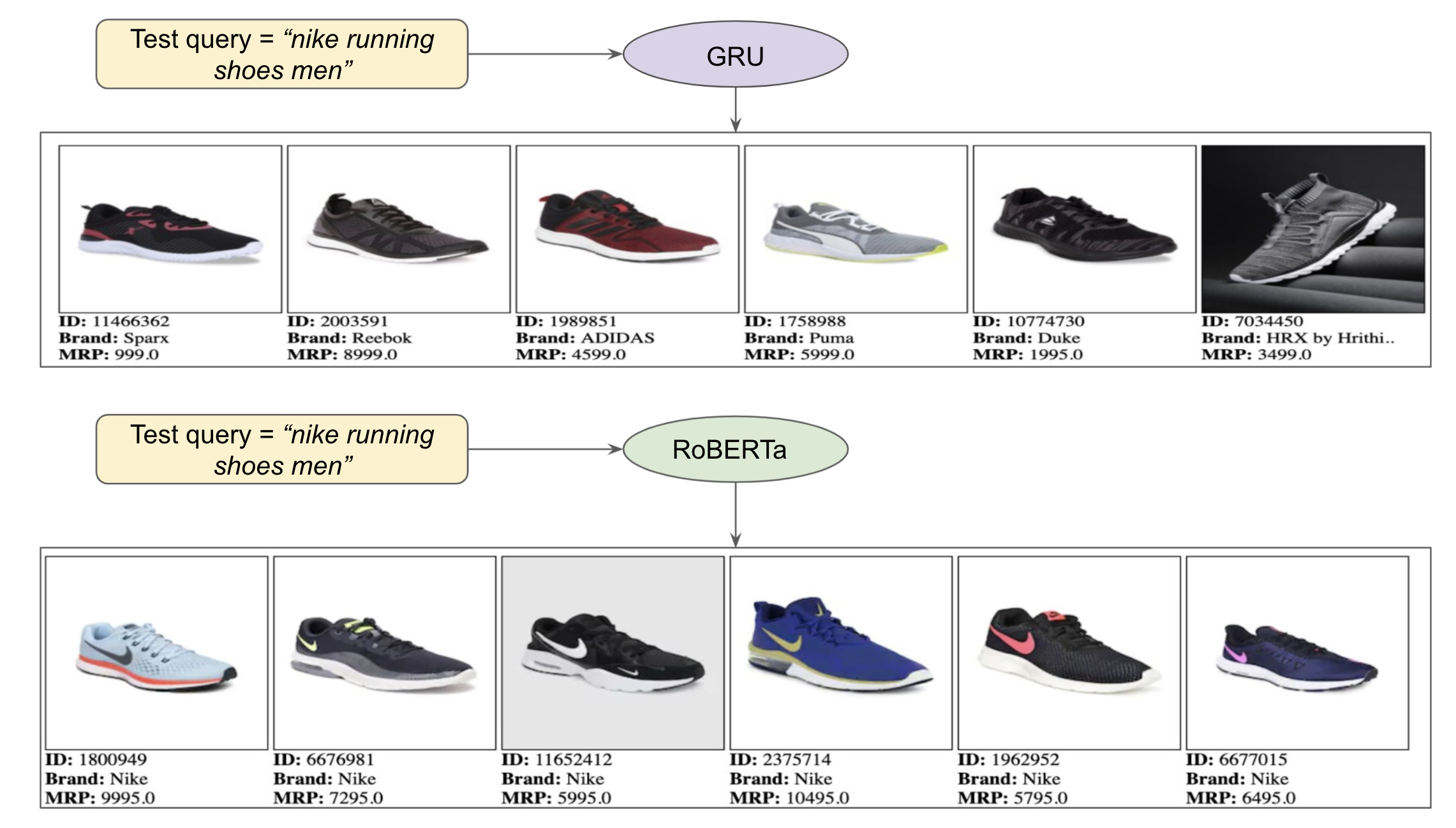}
  \caption{Comparison of retrieved product results for test query: \textit{`nike running shoes men'}}
  \Description{}
  \label{fig:ranking_visualization-3}
\end{figure}

\begin{figure*}[!h]
  \centering
  \includegraphics[scale = 0.38]{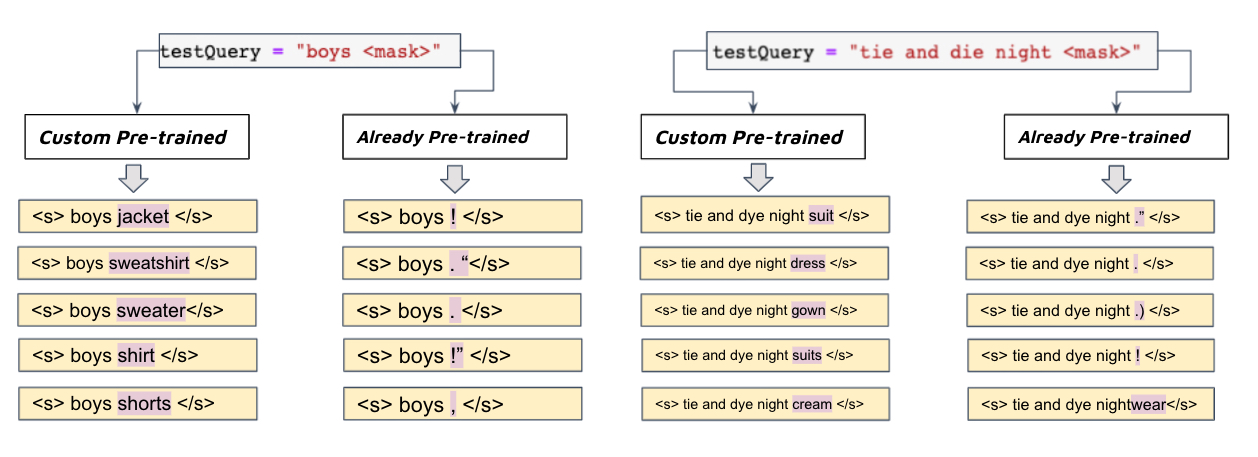}
  \caption{Custom pre-trained vs Already pre-trained RoBERTa for Query Completion }
  \Description{}
  \label{fig:Pre-training Impact}
\end{figure*}

\begin{figure}[!h]
  \centering
  \includegraphics[scale = 0.32]{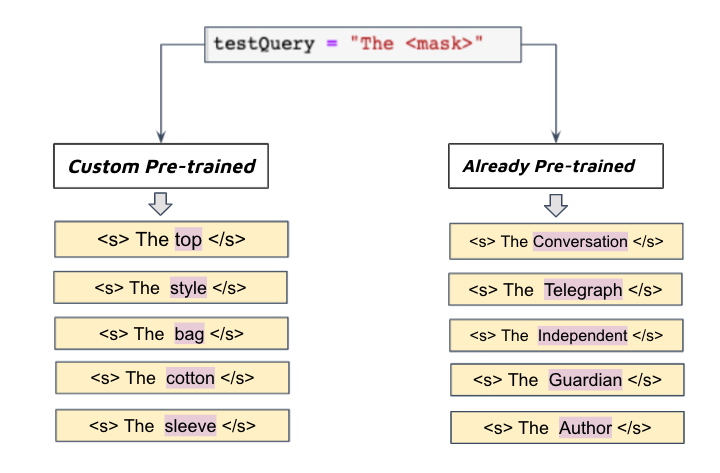}
  \caption{Pre-training impact with respect to the word \textit{`The'}}
  \Description{}
  \label{fig:pre-training impact 2}
\end{figure}

In Figure \ref{fig:ranking_visualization-2} and \ref{fig:ranking_visualization-3}, we show comparison of retrieved product results\footnote{Due to the space limitation, we show less number of retrieved results} for some test queries between GRU and RoBERTa models. For these test queries, we will first pass them through the model to obtain their embeddings. These embeddings are then used to retrieve the products from the product index created using product embeddings obtained from GRU and RoBERTa separately.\\ 
Figure \ref{fig:ranking_visualization-2} show the results for the test query:\textit{`allen solly turtle neck sweatshirts'}. The retrieved products from RoBERTa model clearly indicate the intent of the query, i.e., \textbf{`allen solly'}, \textbf{`sweatshirt'} with \textbf{`turtle neck'}, still some of the results have different neck. Whereas GRU return the products that are \textbf{`sweatshirt'} but from different brands and neck pattern.\\
Figure \ref{fig:ranking_visualization-3} show the results for the test query: \textit{`nike running shoes men'}, and the retrieved products from RoBERTa clearly captures the intent of the query by showing the men running shoes from the \textbf{brand:Nike}. Whereas the retrieved products from GRU model are \textbf{`running shoes'} but from other brands.

Figure \ref{fig:Pre-training Impact} show the visualization for token prediction task with respect to different queries. As we see from the visualizations, after pre-training the RoBERTa model over Fashion corpus, it has a good understanding of the fashion text and generate valid predictions for masked tokens in the queries. However, the already pre-trained RoBERTa model does not generate good predictions and fails to understand the context in some cases. For the test query: \textit{\textbf{`boys <mask>'}}, the custom pre-trained RoBERTa model(pre-trained over the Fashion Corpus) give valid predictions for the <mask> token like \textbf{`jacket', `sweatshirt', `sweater', `shirt', `shorts' etc}. On the other hand the already pre-trained RoBERTa model give garbage predictions in the form of different characters like \textbf{`!', `.'} etc. For one more example test query: \textbf{`tie and dye night <mask>'}, the custom pre-trained model give valid predictions like \textbf{`suit', `dress', `gown', `cream'} etc. But already pre-trained RoBERTa model again give random tokens as predictions. In order to further assess these two RoBERTa model, we give the test query as \textbf{`The <mask>'} as shown in Figure \ref{fig:pre-training impact 2}. The custom pre-trained model give predictions that are coherent with respect to Fashion like \textbf{`top', `style', `bag', `sleeve'} etc. The already pre-trained model give predictions that are coherent with respect to english corpus over which it is pre-trained like \textbf{`Conversation', `Telegraph', `Author'} etc. 
The already pre-trained model is pre-trained over BookCorpus(~16GB)\cite{moviebook}, CC-NEWs(76GB\cite{NewsDataset}), OpenWebText(38GB)\cite{OpenWebText}, Stories(31GB)\cite{storiesDataset} etc which explains the incoherent predictions from this model with respect to Fashion e-commerce. We next highlight the important conclusions and also the future work that can be done to improve ranking and retrieval.
\section{Conclusion \& Future Work}
\label{sec:conclusion}
In this paper we proposed different neural models to learn the representation of queries and products in the latent space. These low dimension query and product representations can be used to solve various problems in the fashion e-commerce domain. We proposed single and multi-layer GRU based baseline models that can be optimized for Triplet Loss using our click-stream data. We also showed how a RoBERTa model after pre-training on Fashion corpus can be again fine-tuned for Triplet loss optimization. Each of the proposed model is trained in two different settings, i.e., first with only Query-Product Data and then augmenting Query-Product data with Product-Product data. In order to compare these models and the representations learned by them, we take the product ranking and retrieval task. Our experiments showed that RoBERTa model that is fine-tuned using only Query-Product data outperformed other proposed models with an MRR of \textbf{48\%}, MAP of \textbf{70.9\%} and NDCG of \textbf{81.1\%} for product ranking task. This model also gave a precision@50 of \textbf{4.5\%} and Recall@50 of \textbf{21.1\%} for product retrieval task. The embeddings learned can also be used to train downstream models for ranking optimization. RoBERTa model acts as a strong baseline which can be directly fine-tuned for different ranking tasks and also leverage the transfer learning due to pre-training over Fashion corpus. Directly fine-tuning the RoBERTa model for ranking optimization is an interesting future work. It will be interesting to compare different transformer based architectures for different ranking and relevance tasks.


\bibliographystyle{ACM-Reference-Format}
\bibliography{main_sigirEcom.bib}
\end{document}